\documentclass[conference]{IEEEtran}
\IEEEoverridecommandlockouts
\usepackage{cite}
\usepackage{amsmath,amssymb,amsfonts}
\usepackage{algorithmic}
\usepackage{graphicx}
\usepackage{textcomp}
\usepackage{xcolor}

\usepackage{xspace}
\usepackage{listings} 
\usepackage{hyperref}
\usepackage{multirow}
\usepackage{soul}
\usepackage{booktabs} % For formal tables
\usepackage{setspace}
\usepackage{enumitem}

\newcommand{\techniqueName}{FILO\xspace}

\newcommand{\RQA}{Are the suspicious invocation blocks capturing information about the symptoms of failures?\xspace}
\newcommand{\RQB}{How well the ranking returned by \techniqueName identifies the method that must be modified to implement the fix?\xspace}
\newcommand{\RQC}{How does \techniqueName compare to both naive trace analysis and SBFL techniques?\xspace}

\newcommand{\urlArtefact}{\url{https://gitlab.com/learnERC/filo}}
\newenvironment{empirical}{\color{black}}{\color{black}}

\def\BibTeX{{\rm B\kern-.05em{\sc i\kern-.025em b}\kern-.08em
    T\kern-.1667em\lower.7ex\hbox{E}\kern-.125emX}}
\begin{document}

\title{\techniqueName: FIx-LOcus Recommendation for Problems Caused by Android Framework Upgrade}

\author{
\IEEEauthorblockN{Marco Mobilio}
\IEEEauthorblockA{\textit{Dept. of Informatics, Systems and Communication} \\
\textit{University of Milano - Bicocca}\\
Milan, Italy \\
marco.mobilio@unimib.it}
\and
\IEEEauthorblockN{Oliviero Riganelli}
\IEEEauthorblockA{\textit{Dept. of Informatics, Systems and Communication} \\
\textit{University of Milano - Bicocca}\\
Milan, Italy \\
oliviero.riganelli@unimib.it}
\and
\IEEEauthorblockN{Daniela Micucci}
\IEEEauthorblockA{\textit{Dept. of Informatics, Systems and Communication} \\
\textit{University of Milano - Bicocca}\\
Milan, Italy \\
daniela.micucci@unimib.it}
\and
\IEEEauthorblockN{Leonardo Mariani}
\IEEEauthorblockA{\textit{Dept. of Informatics, Systems and Communication} \\
\textit{University of Milano - Bicocca}\\
Milan, Italy \\
leonardo.mariani@unimib.it}
}

\maketitle

\begin{abstract}
Dealing with the evolution of operating systems is challenging for developers of mobile apps, who have to deal with frequent upgrades that often include backward \emph{incompatible} changes of the underlying API framework. As a consequence of framework upgrades, apps may show misbehaviours and unexpected crashes once executed within an evolved environment.

Identifying the portion of the app that must be modified to correctly execute on a newly released operating system can be challenging. Although incompatibilities are visibile at the level of the interactions between the app and its execution environment, the actual methods to be changed are often located in classes that do not directly interact with any external element.

To facilitate debugging activities for problems introduced by \emph{backward incompatible upgrades of the operating system}, this paper presents \techniqueName, a technique that can recommend the \emph{method} that must be changed to implement the fix from the analysis of a single failing execution. \techniqueName can also select \emph{key symptomatic anomalous events} that can help the developer understanding the reason of the failure and facilitate the implementation of the fix.
% only, in contrast with localization techniques that require the availability and the execution of full test suites. 

Our evaluation with multiple known compatibility problems introduced by Android upgrades shows that \techniqueName can effectively and efficiently identify the faulty methods in the apps. % changed by the developers to implement the fix.%, outperforming spectrum-based fault localization approaches. %In addition, \techniqueName reports anomalous events observed during the failure that can be useful to implement the fix. 
\end{abstract}


\section{Introduction}

To continuously release new and advanced features that exploit the latest hardware and software upgrades, the operating systems of mobile devices must evolve at a dramatic speed. For instance, the Android API framework evolves at the average rate of 115 API updates per month~\cite{McDonnell:APIStability:ICSM:2013} and a new release is produced every two months in average~\cite{AndroidVersionHistory}. Such a fast \emph{evolution is not problem-free}. For example, in their study Wei et al. found that \emph{more than one third} of the compatibility issues affecting popular Android apps are due to API evolution~\cite{Wei:AndroidFragmentation:ASE:2016}. Note that these problems rarely consist of faults in the framework, but they rather consist of \emph{backward incompatible changes} that require the apps to be fixed to run correctly. This is also confirmed in the study by Mostafa et al. who found that the large majority of backward compatibility problems are fixed in the client code of the apps~\cite{Mostafa:BackwardIncompatibilities:ISSTA:2017}.

Migrating an app to a new API can be \emph{painful}. Once a problem in the app is discovered, developers have to investigate the behaviour of the app to understand the cause of the problem, identify a suitable location for the fix, and implement it. This whole process is demanding and makes developers reluctant to adopt new APIs. For instance, McDonnell et al.~\cite{McDonnell:APIStability:ICSM:2013} reported an average migration time of 16 months, in contrast with an API release interval of few months only. %While apps are under migration, developers have been reported to spend significant time in Q\&A systems like Stack Overflow looking for information about API changes, especially for the API methods with a modified behaviour~\cite{Linares-Vasquez:StackOverflowDiscussions:ICPC:2014}. This further stresses the fact that the 

Simplifying the migration process is thus extremely important to let developers quickly adapt their apps to newly released operating systems. In this paper, we focus on the challenge of assisting the problem resolution task by \emph{automating} the \emph{identification} of the code region that must be modified to fix an app that is incompatible with a newer version of the underlying framework. This can be seen as an instance of spectrum-based fault localization (SBFL)~\cite{Abreu:2007,Jones:Tarantula:ASE:2005}, but contrarily to SBFL that requires a full test suite with passing and failing test cases, our approach, namely \emph{\techniqueName}, requires only a \emph{single} failed test case to be applied. This has three important benefits: (i) it is applicable to the many cases where an extensive unit test suite is not available, (ii) it can be straightforwardly applied to those cases where the failure is exposed with a system-level interaction, such as an automatic system test derived from a bug report entered by a user, and (iii) it is extremely cheap to execute since it avoids the execution of large test suites. 

In contrast with SBFL techniques that can only localize suspicious code regions, \techniqueName also isolates information about the \emph{anomalous events} that are the consequence of the incompatibility between the app and the newly released framework, providing further information potentially useful to the developers to understand the failure and implement a proper fix.

The intuition behind \techniqueName is twofold:
%\begin{itemize}[leftmargin=*]

%\smallskip
\noindent \emph{Interactions between the framework and the app are likely to contain an evidence of the failure}: Since the incompatibility is between an app and its API framework, the problem should intuitively be visible by observing their interactions (i.e., calls from the app to the framework, and vice versa). The comparison of the interactions observed when the app interacts with the compatible (older) and the incompatible (newer) versions of the framework can be used to identify \emph{suspicious interactions} that are in turn useful to identify the \emph{part of the behavior of the app that must be modified} to obtain the fix.

\smallskip
\noindent \emph{The method that must be fixed is likely responsible for a large and coherent set of suspicious interactions}: The detailed analysis of the full failed execution, not only limited to interactions between the app and its framework, can reveal the methods \emph{internal to the app} that control the execution of a significantly \emph{large} and \emph{coherent} set of suspicious interactions. These methods should include the one that must be changed to fix the program. Based on this strategy, \techniqueName generates a ranking that can be exploited by the developers to identify the fix location. Each method can also be associated with the \emph{suspicious interactions under its influence} to provide insights about the rationale of the selection.

\smallskip
%\end{itemize} 

We empirically assessed \techniqueName with multiple incompatibilities between Android apps and various versions of the Android framework. \begin{empirical}Results show that \techniqueName can efficiently identify the method where the fix should be implemented from the analysis of a single failing test case: it ranked the method that must be modified in the top 5 positions in the large majority of the cases, with several cases where the method occurred at the top of the ranking. %, and no case where the target method occurred below the tenth position.
We also compared  \techniqueName to SBFL, confirming its higher effectiveness in addition to its higher applicability.\end{empirical}

In a nutshell, the main contributions of this paper are:
\begin{itemize}[leftmargin=*]
\item \techniqueName, a technique that can address incompatibilities between an app and an updated framework by producing a ranked list of suspicious methods that must be changed, associated with supporting evidence about their selection, from a single system-level failed execution;
\item the empirical evidence that \techniqueName can operate efficiently and effectively;
\item a freely available implementation of our tool and a replication package (\urlArtefact) that can be used to replicate the results reported in  the paper.
\end{itemize}

The rest of the paper is organized as follows. Section~\ref{sec:example} presents a running example that is used throughout the paper to illustrate our approach. Sections~\ref{sec:upgradeHealer} and~\ref{sec:evaluationLeo} describe and empirically evaluate \techniqueName, respectively. Section~\ref{sec:relatedWork} discusses related work. Finally, Section~\ref{sec:conclusions} provides finale remarks.

%The Android system, like many modern software systems, is designed following a model in which the features of the system are accessible via an Application Programming Interface (API) that provides an interface between the applications and the operating system. While this model makes it easy to use the system's hardware and software components, any evolution of the API can lead to compatibility issues that can affect the proper functioning of apps \cite{Wei:AndroidFragmentation:ASE:2016,Mostafa:BackwardIncompatibilities:ISSTA:2017,Li:CompatibilityIssues:ISSTA:2018}.

\section{Running Example} \label{sec:example}
%\todo{Sistemare i Listings}

In this section, we describe an actual Android app that suffers from an upgrade issue common to many others Android apps. The issue is due to the backward incompatible change implemented between the Android API 22 and the Android API 23 that forces apps to explicitly request for a permission when accessing resources for the first time. The opensource \emph{Good Weather} application~\cite{GoodWeather} is one of the applications affected by this backward incompatibility problem, due to an incomplete implementation of support to API 23. The application exploits the location services to produce a weather forecast about the current location of the user. This application has been installed by more than 5,000 Android users according to Google Play. We used this app both in the empirical evaluation and in Section~\ref{sec:upgradeHealer} to illustrate how \techniqueName works. 

%In the empirical evaluation we show how our approach has been effective with a range of diverse incompatibilities, including api methods with a changed semantics and changes to api interaction protocols.

The  method whose implementation is incompatible with API 23 can be obtained from Listing~\ref{lst:gwListing3} by ignoring the red code, which is the code added to obtain the fix, and including the code with the strikethrough font, which is the code removed to obtain the fix. In the faulty implementation, when the {\footnotesize \texttt{gpsRequestLocation}} method is invoked, the permissions to access to the fine and coarse grained locations are checked. If the permissions are granted (as it happens with API 22), the location is regularly updated executing the method {\footnotesize\texttt{requestLocationUpdates()}}. Otherwise, if the permissions are not granted (as it happens with API 23), the method returns without updating the location, resulting in the application hanging forever. The hang is due to the code responsible for the elimination of the progress bar (not shown in the listing) that is executed only once the location has been updated.

%    <@\textcolor{red}{int fineLocationPermission = ContextCompat.checkSelfPermission(MainActivity.this, Manifest.permission.ACCESS\_FINE\_LOCATION);}@>
  %       <@\textcolor{red}{if (fineLocationPermission != PackageManager.PERMISSION\_GRANTED) \{}@>
  %           <@\textcolor{red}{requestLocationPermission();}@>
 %        <@\textcolor{red}{\}}@>

\begin{scriptsize}
\setstretch{0.8} %% RIDUZIONE INTERLINEA
\begin{lstlisting}[language=Java, caption= The Fix for the Good Weather application.,  breaklines=true, label=lst:gwListing3]
public boolean onOptionsItemSelected(MenuItem item) {
  switch (item.getItemId()) {  
    <@\textcolor{black}{...}@>
    case R.id.main_menu_detect_location:
      <@\textcolor{red}{requestLocation();}@>
      <@\st{gpsRequestLocation();}@>      <@\textcolor{black}{...}@>
    }
    <@\textcolor{black}{...}@>
  }
  
  <@\textcolor{red}{private void requestLocation() \{}@>
      <@\textcolor{red}{...}@>
      <@\textcolor{red}{detectLocation();}@>
      <@\textcolor{red}{...}@>
  <@\textcolor{red}{\}}@>
  
  <@\textcolor{red}{private void detectLocation() \{}@>
    <@\textcolor{red}{...}@>
    <@\textcolor{red}{gpsRequestLocation();}@>
    <@\textcolor{red}{...}@>
  <@\textcolor{red}{\}}@>
  
  public void gpsRequestLocation() {
    <@\textcolor{red}{if (checkSelfPermission(this, ACCESS\_FINE\_LOCATION) == PERMISSION\_GRANTED) \{}@>
    <@\st{if (VERSION.SDK\_INT $>=$ VERSION\_CODES.M)\{}@>
       <@\st{if (checkSelfPermission(this, ACCESS\_FINE\_LOCATION) != PERMISSION\_GRANTED \&\& checkSelfPermission(this, ACCESS\_COARSE\_LOCATION) != PERMISSION\_GRANTED)\{}@>
            	<@\st{return;}@>
       <@\st{\}}@>
       locationManager.requestLocationUpdates(LocationManager.GPS_PROVIDER, 0, 0, mLocationListener);
    }
  }
\end{lstlisting}
\end{scriptsize}

In order to fix the program it is necessary to ask the user to grant the access to the location information before invoking the {\footnotesize\texttt{requestLocationUpdates}} method. The developers obtained the fix by modifying the {\footnotesize\texttt{onOptionsItemSelected}} and the {\footnotesize\texttt{gpsRequestLocation}} methods, making them to invoke new methods designed to acquire the required permissions. 
%In addition, the\linebreak \texttt{onRequestPermissionsResult()} callback must be implemented so that the location can be updated just after the user has granted the permissions.
A proper analysis of a failing interaction with the app should report the {\footnotesize\texttt{gpsRequestLocation}} and the {\footnotesize\texttt{onOptionsItemSelected}} methods as the methods to be modified to obtain the fix.% Of course the newly added methods cannot be reported, but reporting the first method, which checks the permissions for the access to the localization service, is likely to be already enough to let the developer understand that the problem is about the access to the localization service. 

In this case, \techniqueName successfully reported the {\footnotesize\texttt{gpsRequestLocation}} and {\footnotesize\texttt{onOptionsItemSelected}} methods at the top of the ranking. In addition \techniqueName can isolate and report anomalous interactions that happened in the failed execution about both permission checking and access to the location service to the developers. These anomalous events are well representative of what the problem is, that is, the app lacks the permission to access the location service.

\begin{figure*}[tb]
  \centering
  \includegraphics[width=0.8\textwidth] {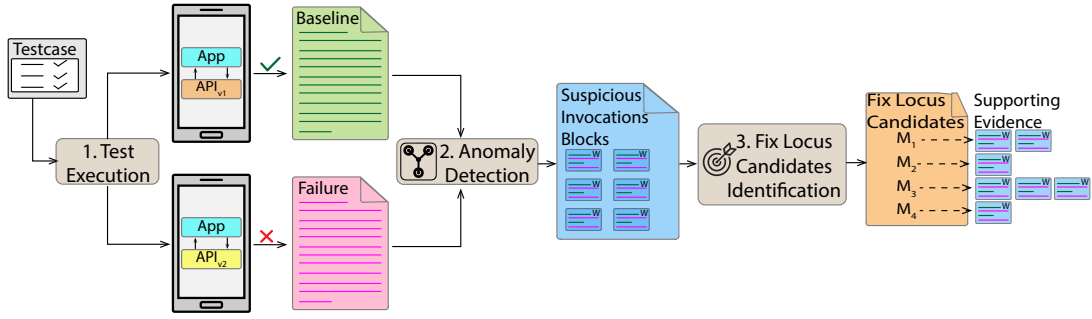}
  \caption{Overview of the \techniqueName technique.}
  \label{fig:overview}
\end{figure*}

Finally, note that the failure per se is not explicative of the fact that the problem is about permissions: the user can only see the hang and the app does not log any error message. This is why the output produced by \techniqueName can be extremely helpful to quickly fix this app.

\section{\techniqueName}
\label{sec:upgradeHealer}

The purpose of \techniqueName is to automatically recommend the likely locations of fixes for problems caused by upgrades of the Android framework to developers. \techniqueName requires three inputs: an \emph{automatic test case}, which is used to reproduce the problem, and the access to two Android environments, one running the app with the \emph{compatible API} and the other running the app with the \emph{incompatible API}. %Both physical and emulated environments can be used \leo{eliminare frase se non risolviamo i problemi con il device fisico}. 
The output produced by \techniqueName is a ranked list of methods corresponding to the possible fix locations, associated with a set of suspicious interactions that motivate the ranking and that can be used by the developer to investigate the problem and implement a fix.

%. In addition, \techniqueName outputs a set of suspicious interactions between the app and the framework that help developers understanding why the app has failed, easing the implementation of the fixes.

%a tool to aid identifying issues in their applications generated by the upgrade of the Android Framework. Within this context we identified three types of issues:
%\begin{enumerate}
%	\item Framework issues introduced by the upgrade that affects the application.
%	\item Framework intended changes of behaviour that affect the application.
%	\item Application issues that did not occur before the Framework upgrade, but happens with the new Framework version.
%\end{enumerate}
%
%All of these kind of issues, may result in both functional and non-functional misbehaviours in third party applications. The expected output of the \techniqueName is twofold: firstly it tries to identify a set of \textit{Suspicious Invocations} (SI), where a SI is defined as a method invocation that is related to the bug. Secondly, it extracts a ranking of \textit{Fix Locus Candidates} (FLC), where a FLC is a method that should be modified in order to fix the bug.

\techniqueName works in three main phases as shown in \figurename~\ref{fig:overview}: the \emph{Test Execution} phase runs the test case that reproduces the failure and collects the interactions between the app and the framework from the two available environments; the \emph{Anomaly Detection} phase identifies blocks with suspicious interactions by comparing the collected traces; the \emph{Fix Locus Candidates Identification} phase identifies and ranks the places where a fix is likely to be implemented and associates the corresponding evidence. We describe these three phases in details below.
%The rest of this section describes these three phases in details.

\subsection{Test Execution}
The test execution phase collects the interactions between the app and the framework for both the  \emph{base environment}, which runs the target app with a \emph{compatible} version of the framework API, namely \emph{v1}; and the \emph{upgraded environment}, which runs the same target app with an incompatible version of the framework API, namely \emph{v2}. The collected interactions consist of all the calls to methods of the framework produced by the app, and all the calls to methods of the app produced by the framework. While the rest of the calls, that is, calls internal to the framework and calls internal to the app, are ignored. This is because incompatibilities must be observable while looking at the interactions between the app and its framework. %, and thus from the api and callback calls produced during a failure. 

More formally, a \emph{framework method} is a method defined in the Android framework or in a standard library (e.g., {\footnotesize\texttt{java.*}}); an \emph{application method} is a method implemented in the app. An \emph{API call} is a call to a framework method originated by an application method. Vice versa a \emph{callback} is a call to a method of the app originated by a framework method. All the other cases are internal method calls that are ignored in this step (note that calls internal to the app are on the contrary relevant to the third step for the identification of the fix locus). %Figure \ref{fig:Invocations} visually illustrates all the cases. 

The trace files recorded by \techniqueName only include API calls and callbacks. We refer to the union of the API calls and callbacks as the \emph{boundary calls}. To collect this information \techniqueName instruments the app and executes the automatic test case on both the base and upgraded environments.

%\begin{figure}[htb]
%  \centering
%  \includegraphics {"Invocations"}
%  \caption{The different types of calls.}
%  \label{fig:Invocations}
%\end{figure}

The output of this phase consists of two traces containing an interleaved sequence of API calls and callbacks. The \emph{baseline trace} is obtained by running the test case within the base environment and represents how the app and the framework interact when the execution is correct. The \emph{failure trace} is obtained by running the test case within the upgraded environment and represents how the app and the upgraded framework interact  when the app fails. 

\begin{scriptsize}
\begin{lstlisting}[language=Java, caption= Excerpt of the Good Weather baseline trace.,  breaklines=true, label=lst:traceListing]
<@\emph{MainActivity.onCreate()\#b}@>
AppCompatActivity.onCreate()#b
AppCompatActivity.onCreate()#e
AppCompatActivity.getSupportActionBar()#b
AppCompatActivity.getSupportActionBar()#e
AppCompatActivity.setContentView(...)#b
AppCompatActivity.setContentView(...)#e
AppCompatActivity.findViewById(...)#b
AppCompatActivity.findViewById(), returnValue:...#e
AppCompatActivity.setSupportActionBar(...)#b
AppCompatActivity.setSupportActionBar(...)#e
<@\emph{MainActivity.onCreate()\#e}@>
\end{lstlisting}
\end{scriptsize}

An excerpt of a baseline trace is shown in Listing~\ref{lst:traceListing}. Note that \techniqueName traces when the execution of every boundary call both starts and ends (marked with \texttt{\#b} and \texttt{\#e} in the sample trace). If a boundary call produces a return value, \techniqueName also records it in the trace (such as for {\footnotesize\texttt{findViewById()}} in the sample file): if the return value is of primitive type \techniqueName directly records it, if the return value is non-primitive \techniqueName records the value returned by method {\footnotesize\texttt{toString()}} if overridden, otherwise \techniqueName only records the dynamic type of the return value. In the example trace we indicate callbacks in italics: only the call to\linebreak {\footnotesize\texttt{MainActivity.onCreate()}} is a callback since the method is invoked from a method of the framework (not shown in the trace). All the other calls are API calls, that is, they are calls to the framework ({\footnotesize\texttt{AppCompatActivity}} is a class of the framework) generated by the app ({\footnotesize\texttt{MainActivity.onCreate()}} is implemented in the app).

Executing \emph{exactly} the same test in both environments and restricting the observations to boundary calls reduce the risk of having incidental differences in the traces, which implies that the large majority of the observed differences are relevant to the problem under analysis. If non-deterministic interactions are present in the traces, they can be filtered out by executing the same test multiple times and eliminating the changing portion of the trace from the analysis. We however never observed such case in our evaluation.

%In principle, it is still possible to observe non-deterministic behaviors in the traces, but in our evaluation we never observed such cases. If necessary, the non-deterministic interactions between the app and the framework can be filtered out by executing the same test multiple times and eliminating the non-deterministic portion of the trace. 

%\review{
%I do not agree that the differences between two traces can only be due to changes to the underlying framework and their impact on the app. What about non-determinism due to asynchronous taks and concurrency? What about transient behaviors due to changes to the environment (e.g., the network)? 
%1. Are the differences between two traces necessarily due to changes to the underlying framework? What about non-determinism or transient behaviors?
%}
%\marco{lanciare piu' volte il test sulle due versioni potrebbe effettivamente contribuire a identificare differenze non legate al bug (che si ripetono in esecuzioni diverse sulla stessa API) che potrebbero quindi essere escluse}
Note that although the scope of the observation is limited to boundary calls, the size of the traces might be significant. For instance, the baseline trace collected for the running example includes 14,245 method calls and has a size of  $\sim$41MB, while the size of the traces in the experiments \begin{empirical}reached 617,427 method calls and $\sim$1.8GB for the MapDemo application~\cite{MapDemo}\end{empirical}. %\leo{Number of events and size of the longest trace}.
%\marco{GoodWeather e' rimasta la piu' grande in termini di tracce non filtrate, se consideriamo le tracce di partenza raccolte dal nostro Logger verrebbe una cosa cosi': For instance, the Baseline trace collected for the running example includes 14245 method calls and weights about \textasciitilde 41MB, while the size of the traces in the experiments reached 617427 method calls and \textasciitilde 1.8GB in the case of MapDemo.}

%Identifying the incompatibility in the trace and tracing it back to the method that must be modified to fix the program is thus still challenging. 

\subsection{Anomaly Detection} \label{subsec:tech_anomaly}
The anomaly detection phase compares the two traces produced in the test execution phase and isolates a number of invocation blocks that look suspicious. An \emph{invocation block} is a contiguous sequence of boundary calls extracted from the traces generated during test execution. More formally, given a recorded trace $e_1, \ldots e_n$, an invocation block is any subsequence $e_i, e_{i+1}, \ldots e_{i+k}$, s.t., $i\geq 1$ and $i+k \leq n$.

The anomaly detection phase identifies, groups, and assigns weights to the differences between the baseline trace and the failure trace. Since the two traces show the boundary calls collected while running exactly the same test case, most of the differences are likely due to changes in the underlying framework and their impact on the app, specifically in the case of the failure. \techniqueName identifies these differences by running \emph{diff}~\cite{LinuxDiff}, which returns the invocation blocks in the failure trace with no counterpart in the baseline trace. Since these blocks represent differences that may characterize the failure, we refer to them as the \emph{Suspicious Invocation Blocks} (SIBs). 

The set of differences contained in the SIBs can still be large. Indeed a different behavior of the framework may produce several differences in the return values of the API calls and on the generated callbacks. As a consequence the app may react differently than with the older version of the framework and produce unexpected boundary calls. For instance, in the running example there are thousands of differences between the baseline and failure traces.

To analyze the content of the SIBs, \techniqueName performs two operations: it associates the block with its first boundary call, since that call is likely to be the cause of all the differences reported in the block; in addition, considering that spurious interactions not related to the failure are often due to short SIBs of length 1, \techniqueName weights the relevance of each block based on the number of boundary calls that it contains.

%Since differences are often in a stream, that is, one difference can be the immediate consequence of a former difference and so on, we group sequential differences together creating what we called \emph{suspicious invocation blocks}\footnote{When clear from the context we refer to the suspicious invocation blocks simply as the blocks.}, where the first boundary call of the block is used as representative of the full content of the block, since it is the source of all the differences.

%Considering that spurious interactions not related to the failure are often due to short and many times even to suspicious invocation blocks of length 1, we weight the relevance of each block based on the number of boundary calls that it contains.

For example, Listing~\ref{lst:diffTrace} shows an excerpt of a SIB detected for the running example. The block starts with a call to method {\footnotesize\texttt{MainActivity.onLocationChanged}}, which is representative of the entire stream of anomalous interactions that follow. Since the block includes 30 boundary calls, this is also the weight of the block, which is shown on the first row of the listing.

The output of this phase is thus a number of weighted SIBs that \techniqueName uses to identify how and where to change the app to fix the compatibility problem. Note that the anomaly detection step typically identifies several blocks. For instance, the number of SIBs in the running example are 98. However, not all of them have the same relevance based on our heuristics, indeed many of them are short spurious anomalies. For example, 70 SIBs have a weight of $1$ (i.e., they include a single boundary call) and only $23$ SIBs have a weight higher than 3.

Note that the set of SIBs often does not contain the method that should be modified to fix the program but only the methods that misbehave due to the upgrade. For instance, one of the methods that must be modified in the running example is {\footnotesize \texttt{gpsRequestLocation()}} and this method never occurs in the baseline and failure traces. This is confirmed in our empirical evaluation, where for 50\% of the cases the recorded traces do not include the method with the fix. 

%two traces of the boundary methods collected in the Base and Upgraded environments.
%\review{
%Second, is it really the case that the fault-related method calls (i.e., the calls to the methods to to be fixed) are typically (1) not part of the suspicious invocation blocks and (2) on the stack when the block is executed? It would be nice to see some evidence that supports these assumptions, as I can think of cases in which they would not hold
%2. Is there any strong evidence that fault-related method calls are typically not within a suspicious invocation block?
%}

\begin{scriptsize}
\begin{lstlisting}[language=Java, caption= Excerpt of a Suspicious Invocation Block,  breaklines=true, label=lst:diffTrace]
<@\emph{MainActivity.onLocationChanged(...)\#b}@> Weight 30
android.app.Dialog.cancel()#b
android.app.Dialog.cancel#e
android.location.Location.getLatitude()#b
android.location.Location.getLatitude, returnValue: ...#e
...
\end{lstlisting}
\end{scriptsize}

\subsection{Fix Locus Candidates Identification }\label{subsec:fixLocusCandidates}

The \textit{fix locus candidates identification} phase is the last phase of the \techniqueName technique. It takes the set of weighted SIBs as input and produces a ranked list of methods that represent the locations where the fix should be likely implemented as output. Each method in the ranking is associated with supporting evidence that consists of the set of SIBs that might be affected by the method. Intuitively, the suspicious invocations associated with a method represent the symptoms of the failure that can be removed by implementing a proper fix in the method. This information can be useful for the developers who can benefit from contextual information and insights about the failure, in addition to information about the location where a fix should be implemented, when working on the app to produce a fix.

%The method that must be modified is usually ranked at the very top positions. The suspicious invocation blocks assigned with high weights provide additional information to the developer for the design of a fix.

Since the fixes must be often implemented in methods that are not part of the SIBs, which in fact represent the symptoms of failures and not their causes, \techniqueName considers any method executed in the failing execution as a potential target for the fix. In particular, \techniqueName creates the \emph{failure call tree}, which is a tree that includes all the methods present in the stack trace at the time each SIB has been detected (this information is collected when executing the failing test). Nodes represent the invoked methods and direct edges represent calls between methods. %The root of the failure call tree is thus the \texttt{ZygoteInit.main} method and the leafs are the suspicious invocation blocks returned by the anomaly detection phase.
More formally, for each SIB $\textit{sib}_i$ with weight $w_i$ and representative call $c_i$, \techniqueName collects its stacktrace $\langle m_1, \ldots, m_{n_i}, c_i \rangle$, where $m_1$ is always the method {\footnotesize\texttt{ZygoteInit.main}}. The failure call tree is a triple $(N, E, r)$, where $N= \{m_1, \ldots, m_{n_i}, c_i | \forall \textit{sib}_i\}$ is the set of nodes, $E= \{(m_{n_i},c_i), (m_i, m_{i+1}), i=1, \ldots n_i-1 | \forall \textit{sib}_i\}$ is the set of edges, and $r$ = {\footnotesize\texttt{ZygoteInit.main}} is the root of the tree.

Note that the root of the failure call tree is the {\footnotesize\texttt{ZygoteInit.main}} method, which is the initial method that handles the forking of every application launched in Android. The leafs are the representative methods of the SIBs, returned by the anomaly detection phase. Each leaf node has a weight corresponding to the weight of the SIB that originated the node. Leaf nodes with high weights are more relevant to the analyzed failure than the other leaf nodes. 

  \techniqueName scores each node of the tree, that is, each method call, based on its degree of influence on the SIBs, represented by the weighted leaf nodes of the failure call tree. Intuitively a high score indicates a method that can directly affect the execution of several blocks of non-trivial weight. This score is the \emph{suspiciousness} of the method and is used to produce the final ranking. More formally, we compute the suspiciousness of a method $m$ as a linear combination of two attributes based on the following formula:
%\begin{small}
%\begin{displaymath}
%\textit{Susp}(m) = k_1  \textit{ImpBlocks}(m) + k_2  \textit{Depth}(m) + k_3 \textit{Children}(m)
%\end{displaymath}
%\end{small}
\begin{small}
\begin{displaymath}
\textit{Susp}(m) = k_1  \textit{ImpBlocks}(m) + k_2  \textit{Depth}(m) 
\end{displaymath}
\end{small}
The sum $k_1+k_2$ is 1. Each attribute measures a characteristic that should be taken into account in the attempt of identifying the method that is likely the locus of the fix and its value ranges between 0 and 1. We discuss these two attributes below.

%\begin{figure*}[htb]
%  \centering
%  \includegraphics [width = \textwidth]{"GoodWeatherGraph"}
%  \caption{Graph created for the GoodWeather Application.}
%  \label{fig:goodWeatherGraph}
% % \vspace{-0.5cm}
%\end{figure*}

\smallskip

%\begin{itemize}[leftmargin=*]
%\item 
\emph{ImpBlocks(m)} measures the number of SIBs that can be affected by a specific method. It is computed by summing the weights of the SIBs (leaf nodes) that can be reached from $m$ following a path in the failure call tree. The value is normalized with respect to the sum of the weights of all the SIBs in the tree. Note that by definition the root of the failure call tree can affect all the SIBs, thus $\textit{ImpBlocks}(\textit{root} \textit{node})=1$. Similarly a node $sib$ representing a SIB with weight $w$ can only influence itself and thus $\textit{ImpBlocks}(\textit{\textit{sib}})=\frac{w}{W}$, where $W$ is the sum of the weights of all the SIBs in the tree.

%\item  
\smallskip

\emph{Depth(m)} measures the position of the node with respect to the height of the failure call tree, that is, it measures how close a selected method is to the SIBs. The score is normalized with respect to the height of the tree. Thus, $\textit{Depth}(\textit{root} \textit{node})=0$ and $\textit{Depth}(\textit{sib})=1$ for the deepest SIB  $\textit{sib}$.

%\item \emph{\textit{Children}(m)} measures the number of direct children of a node, normalized with respect to the maximum number of children of a node in the failure call tree. This attribute intuitively captures the complexity of a method in terms of the number of operations that it may trigger. 
%\end{itemize}
\smallskip

These two attributes suitably interact one with the other to identify the methods that with the highest probability must be changed to fix the failure. The \emph{ImpBlocks(m)} attribute privileges the choice of nodes at the top of the tree selecting methods that control the execution of a significant  number of SIBs. This intuitively satisfies the criterion that the method that must be changed to implement the fix \emph{must have an impact on a significant number of the SIBs} that have been detected.

The \emph{Depth(m)} attribute privileges the choice of nodes close to the leafs of the tree, discouraging the selection of methods that have been executed too early during the failure. The interaction of this attribute with the \emph{ImpBlocks(m)} attribute favours the selection of \emph{nodes that influence the execution of a cohesive set of SIBs with relevant weights}. In fact, starting from the selection of a leaf node (i.e., a SIB), it is worth selecting an ancestor node, that is, a method executed earlier in the failure, only if the loss in the $\textit{Depth}$ attribute is compensated by the gain in the $\textit{ImpBlocks}$ attribute. In particular, it is convenient to consider methods that can control the execution of a higher number of SIBs only if these blocks have a relevant weight.

Since the failure call tree represents method calls and the same method might be invoked in multiple places, while \techniqueName ranks methods based on their suspiciousness, the final ranking is obtained by considering each method only once using the occurrence with the highest suspiciousness for the ranking. Before returning the ranking to the user, all the methods belonging to the framework are removed because the adaptation must necessarily target the app. In the running example, the resulting ranking is shown in Table~\ref{tbl:gwRankingEx}. Note that the methods that require the fix, that is {\footnotesize \texttt{gpsRequestLocation}} and {\footnotesize\texttt{onOptionsItemSelected}}, are ranked at the top, they are thus the first methods that a developer would inspect.

Since SIBs with just one call, that is, blocks of weight 1, in the large majority of the cases correspond to noisy differences unrelated with the failure, \techniqueName uses only blocks of minimal weight 2, unless all the blocks have weight 1.

Note that in the running example the top ranked method identified by \techniqueName  never occurs in the set of boundary calls and thus a trivial comparison of the log files could not produce a precise recommendation to the developer.

\begin{table}[]
\caption{Ranking returned by \techniqueName.} \label{tbl:gwRankingEx}
\vspace{-0.3cm}

%\resizebox{\textwidth}{!}{%

\resizebox{0.48\textwidth}{!}{%

%\begin{small}
\begin{tabular}{@{}ll@{}}
\toprule
MethodName                                                        & Susp                  \\ \midrule
\textbf{org.asdtm.goodweather.MainActivity.gpsRequestLocation}    & \textbf{0.72} \\
\textbf{org.asdtm.goodweather.MainActivity.onOptionsItemSelected}          & \textbf{0.69}          \\
org.asdtm.goodweather.MainActivity\$1.onLocationChanged           & 0.53          \\
org.asdtm.goodweather.MainActivity.preLoadWeather                 & 0.45          \\
org.asdtm.goodweather.MainActivity.onResume                       & 0.43          \\
org.asdtm.goodweather.BaseActivity.configureNavView               & 0.42          \\
org.asdtm.goodweather.MainActivity.onPause                        & 0.39          \\
org.asdtm.goodweather.BaseActivity.setupNavDrawer                 & 0.39          \\
org.asdtm.goodweather.BaseActivity.onPostCreate                   & 0.36          \\

%org.asdtm.goodweather.BaseActivity.setContentView                 & 0.34          \\
%org.asdtm.goodweather.BaseActivity.setupNavDrawer                 & 0.33          \\
%org.asdtm.goodweather.MainActivity.onCreate                       & 0.32          \\
%org.asdtm.goodweather.MainActivity.onPause                        & 0.30          \\
%org.asdtm.goodweather.BaseActivity.onPostCreate                   & 0.29          \\ 
\bottomrule
\end{tabular}
%\end{small}
}
\vspace{-0.5cm}
\end{table}

Finally, \techniqueName cannot address faults that occur in methods that are not part of the stacktraces collected when the SIBs are detected. Although this may sometime happen, a failing execution produces several SIBs and thus a number of stacktraces, and consequently a number of methods are added to the failure call tree during the analysis. The possibility that the faulty method is not part of the tree is small, as also confirmed in our evaluation where this happened only in 1 case out of 12. We thus decided to design \techniqueName to be fast and effective, at the cost of potentially missing a small percentage of faults.

\section{Empirical Evaluation} \label{sec:evaluationLeo}

The empirical evaluation answers to three research questions concerning the information captured in the SIBs, the quality of the ranking, and the comparison to \emph{naive trace analysis} and \emph{spectrum based fault localization} (SBFL).

\noindent \textbf{RQ1: \RQA}
This research question investigates if the SIBs really capture the symptoms of the failures and thus can be reliably used to build the rest of the analysis. To answer this research question we compare the content of the SIBs to the set of method calls actually related to the incompatibility between the framework and the app. %, and that would be changed by the fix.

\noindent \textbf{RQ2: \RQB} This research question investigates if \techniqueName is able to produce rankings where the method that must be modified to fix the program occurs at the top positions of the ranking, possibly within the first 10 positions, and ideally within the first 5~\cite{Kochhar:LocalizationExpectation:ISSTA:2016}.
 
\noindent \textbf{RQ3: \RQC} This research question compares \techniqueName to two competing approaches. Naive trace analysis consists of identifying the method that must be changed simply comparing the baseline and failure traces one to the other. This comparison has the objective to assess the importance of the third phase of the technique, which goes beyond the content of the trace files for the localization. We also compare \techniqueName to SBFL techniques, Ochiai~\cite{Abreu:2007} in particular, in terms of the ability to identify the faulty method. Although \techniqueName operates under weaker assumptions than SBFL, for instance \techniqueName does not require the availability of a full test suite with passing test cases to be applied, we compared them to study how their effectiveness relate one to the other.

\smallskip

In addition to the three research questions, we qualitatively discuss the supporting evidence generated by \techniqueName in a discussion section. In particular, we show how the SIBs isolated by \techniqueName for the methods in the ranking may represent a useful support to understand the failure and possibly implement a fix, despite the presence of noisy method calls. 
We conclude the section presenting the threats to the validity of the experiment and discussing the limitations of the approach. 

\subsection{Prototype}

% !TEX root =  Main.tex
%
%
%\begin{table*}[t]
%\centering
%\caption{Subject Apps.}
%\label{tab:apps}
%
%\begin{tabular}{@{}l | l l l l | l l @{}}
%Application      & locs & apk size & \# forks & Incomp. API & Failure & Faulty behavior \\ \midrule
%
%BossTransfer     & 0.4K    & 1.4MB & 43 & 23 & crash when opening the details about items in a list & wrong permission logic\\
%
%FakeGPS          & 1.6K & 1.4MB & 60 & 23 & crash when opening the view to set the fake position & missing permission logic\\
%
%FilePicker          & 1.9K & 1.8MB & 28 & 23 & folders erroneously shown as empty & faulty support to the new api\\
%
%GoodWeather    & 3.9K & 3.2MB & 51 & 23 & hang when refreshing meteo forecast & missing permission logic \\
%
%KanjiFix        	& 3.2K   & 5.6MB & 3 & 21 &  unable to fix Japanese glyph rendering & fonts require a new procedure to be loaded\\
%
%MapDemo        & 0.2K & 2.1MB & 1 & 23 &  crash when retrieving the current position & missing permission logic\\
%
%onInflate        &  0.1K & 0.3MB & 0 & 23 & dataloss after screen rotation &  missing callback   \\
%
%SearchView        & 2.6K & 2.6MB & 352 &  21 & crash on  startup &  faulty support to the new api\\
%
%\bottomrule
%\end{tabular}
%\end{table*}

\begin{table*}[t]
\centering
\caption{Subject Apps.}
\label{tab:apps}
\begin{footnotesize}
\begin{tabular}{@{} l c c c | c | l l@{}}
\toprule
Application      & Ver & Locs  & Inc. API & Failure & Fault \\ \midrule

BossTransfer     & 47 & 1.3K  & 23 & crash when opening the details about items in a list & wrong permission logic*\\

FakeGPS          & 28  & 3.0K & 23 & crash when opening the view to set the fake position & missing permission logic*\\

FilePicker          & 115& 5.8K  & 23 & folders erroneously shown as empty & faulty support to the new api\\

GetBack GPS        & 2133 & 7.0K & 23 & unable to retrieve current position &  missing permission logic\\

GoodWeather    & 745& 9.7K & 23 & hang when refreshing meteo forecast & missing permission logic \\

KanjiFix        	& 46& 1.3K  & 21 &  unable to fix Japanese glyph rendering & fonts require a new procedure to be loaded\\

MapDemo        & 5 & 0.6K  & 23 &  crash when retrieving the current position & missing permission logic*\\

%onInflate        &  23 & dataloss after screen rotation &  missing callback   \\
PoGoIV & 2328 &  19.2K & 24 & unable to perform the auto update & new api requires the use of FileProvider\\
%%"Summarily, file:// is not allowed to attach with Intent anymore or it will throw FileUriExposedException which may cause your app crash immediately called."

PrivacyPolice & 153& 2.5K  & 23 & unable to connect to wifi networks & api methods with changed semantics\\

QuotoGraph        & 289 & 8.4 K & 24 & crash on  startup &  api methods with changed semantics\\

SearchView        & 746 & 6.1K & 21 & crash on  startup &  api methods with changed semantics\\

ToneDef        & 91 & 6.8K & 23 & error message when dialling from the phone contacts list &  missing permission logic*\\

\bottomrule
\end{tabular}
\end{footnotesize}
\vspace{-0.5cm}
\end{table*}

Our prototype implementation consists of two main components: the tracer, which is responsible of recording the traces with boundary calls, and the analyzer, which is responsible of producing the ranking. Since Android represents the largest share of smartphone operative systems (86,8\% of the units shipped in 2018Q3 are Android phones~\cite{AndroidShare}), we implemented these two components targeting the Android ecosystem. Note that, although our implementation is specific to Android, \techniqueName is general and could be implemented also for other ecosystems.  % as well, such as Apple iOS. 

To make our results reproducible by third parties we implemented, for each app, the shortest interaction sequence that reveals the incompatibility between the app and the framework as an automatically executable Appium test case~\cite{Appium}. We packaged and made our prototype implementation, the apps studied in this paper, and the corresponding Appium test cases available online at the following url: \urlArtefact.  

The tracer collects the information about failures by running the Appium test case twice. The first time the tracer exploits the native Android Tracer to obtain the list of executed methods. These methods are then used to configure our monitor implemented as an Xposed~\cite{Xposed} module to selectively instrument the relevant methods only, collecting additional information that cannot be extracted with the Android Tracer, such as the return values, and the stack-trace of the invocations. 
The final trace with boundary calls only is obtained by filtering the collected calls based on the identity of the callers.

The analyzer is a Java component analyzing the traces as discussed in Section~\ref{sec:upgradeHealer}. To configure the values of the two weights $k_1$ and $k_2$, we empirically evaluated multiple combinations with a subset of the apps and we ended up using $k_1=0.25$ and $k_2=0.75$. %we experimented the approach with a selection of four apps (GoodWeather, PoGoIV, PrivacyPolice, and SearchView). 
%We ended up using $k_1=0.25$ and $k_2=0.75$.
\techniqueName demonstrated to be relatively sensitive to the choice of $k_1$ and $k_2$ since non-trivial variations of their values had little impact on the results. Specifically, values in the range $0.01<k_1<0.34$ and $0.66<k_2<0.99$ worsen the ranking by moving the target method only 1 place below in average.

\subsection{Subject Programs}

To evaluate our approach we searched for Android apps that presented issues after a framework upgrade on GitHub. We used keywords such as "after upgrade to Lollipop" for the initial selection. %\cameraReady{While the search strategy may have been simple yielding incomplete results, the selected apps define an interesting benchmark for experimentation since they belong to different domains and present different classes of failures.} 
Since our evaluation requires both the possibility to reproduce the failures and the knowledge of the location of the fix, we filtered out apps where it was impossible to replicate the upgrade problem (e.g., because it was impossible to compile the project or failure reproduction required a specific hardware). When available we used the official fixes, otherwise we implemented the fix ourselves. %Although we found more than 100 issues with our initial selection, the lack of an available fix and the technical difficulties in the compilation and reproduction of the failure scenarios 
We \begin{empirical}ended up with 12 actual upgrade problems\end{empirical} and corresponding Appium test cases that replicate the problem. 
%\review{Another issue is that the test case used to expose the problem caused by the upgrade is likely to play an important role in the success of the technique. However, the paper does not mention at all how the specific test case used for an app was selected. Again, the author should explain how they picked their tests and perform at least an initial sensitivity analysis to understand how much the results depend on the specific test considered.}
%\marco{Most of the apps found on the search do not have a test suite, so we developed tests that highlights the bug for each application, mostly by replicating the steps reported in the issues related tho the bugs.}
Table~\ref{tab:apps} reports information about the apps, the failures, and the corresponding faults. Columns \emph{Application}, \emph{Ver}, and \emph{Locs} indicate the name, the version of the app (specified with the identifier of the commit), and the number of lines of code of the app, respectively. BossTransfer is a game app~\cite{BossTransfer}. FakeGPS is a GPS device simulator~\cite{FakeGPS}. FilePicker is an app for selecting files and folders in a device~\cite{FilePicker}. GetBack GPS is an app for storing the location of points of interests~\cite{GetBackGPS}. GoodWeather is a weather app~\cite{GoodWeather}. KanjiFix is an app to fix Japanese glyph rendering~\cite{KanjiFix}. MapDemo is an app to test the setup of Google Play services~\cite{MapDemo}. PoGoIV is an IV calculator for Pokemon Go~\cite{PoGoIV}. PrivacyPolice is an app that prevents leaking of sensitive information via Wi-Fi networks~\cite{PrivacyPolice}. QuotoGraph is an app to create custom wallpapers~\cite{QuotoGraph}. SearchView is a persistent search and view library in material design~\cite{SearchView}. ToneDef is a tone dialer application~\cite{ToneDef}.% featuring DTMF, bluebox and redbox tone generation~\cite{ToneDef}.
%\review{For how many of the 12 apps, did you have to implement the fix and how may had developer-implemented fixes available? In the latter case, were the fixes spread across multiple commits? Are all the apps actively used and maintained? Please provide data on downloads, commits, etc. 5000 downloads for an app (GoodWeather) is not particularly high. Could you have used apps evaluated in prior studies (e.g., [59])? Some of those apps in fact contain API-incompatibility bugs.}

Column \emph{Inc. API} reports the version of the framework API that is incompatible with the app. In most of the cases it is API 23 since it is also the most used version of the framework~\cite{AndroidDistribution}. 
Column \emph{Failure} provides a short description of the failure caused by the incompatibility with the framework API. 
Finally column \emph{Fault} provides a short description of the fault that causes the failure of the app. The * indicates that we implemented the fix since the official fix was not available. 

Note that some of these faults have been non trivial to debug, requiring  from 1 day to several months to be fixed. Three of the faults  also required multiple commits to be fixed (up to 23 commits in one case). 

\subsection{RQ1: \RQA}

%\review{it is not clear how the metrics were computed. Additionally, what happen with sub-blocks that contain faulty methods?
%}

To answer this research question, we analyzed the content of the SIBs measuring the completeness and soundness of the reported information. To \emph{objectively} identify the set of anomalous boundary calls caused by a fault we exploited the fixed versions of the apps. %In particular, we downloaded from GitHub the version of the app that differs from a faulty app for the fix only. 
We executed both the faulty and the fixed apps on the upgraded framework using the test case that reproduces the problem and collected the boundary calls. We then compared the two trace files. Differences correspond to suppressed or novel method calls introduced by the developer to fix the program. We refer to these calls as the \emph{fault-related method calls} (\emph{frmc}). Ideally, the SIBs should be able to capture these calls, which would be then exploited to rank methods.
 
To measure the \emph{soundness} of the content of the SIBs we compute the percentage of blocks that contain fault-related method calls, that is, $\textit{soundness} = \frac{\# \textit{SIB}\ \textit{with}\ \textit{frmc}}{\# \textit{SIB}}$. We perform this for both all the SIBs and for the blocks with a minimum length of 2, which are the ones used by \techniqueName unless all the blocks have length 1.  
To measure the \emph{completeness} of the content of the SIBs, we compute the percentage of fault-related method calls that are included in the SIBs, that is, $\textit{completeness} = \frac{\# \textit{frmc}\ \textit{in}\ \textit{SIBs}}{\# \textit{frmc}}$. Table~\ref{tab:sib} summarizes the results.

\begin{table}[t]
\caption{Suspicious Invocation Blocks.} \label{tab:sib}
%\begin{tabular}{@{}llll|lll@{}}
\begin{center}
\begin{footnotesize}
\begin{tabular}{@{}lcc | c}
\toprule
 
\multirow{2}{*}{Application}  &  \multicolumn{2}{c|}{Soundness}  & \multirow{2}{*}{Completeness}\\                                                                                                                                                                                                                                                                                                                                        
% \multirow{ 3}{*}{Appl}  & Soundness & pippo  & Completeness\\ 
 %\midrule
 \cmidrule{2-3}
& All SIBs & SIBs ($\textit{length} \geq 2$) &  \\
%& & ($\textit{length} \geq 2$) & Method Calls\\

%            & \multicolumn{2}{c|}{Soundness}  & Completeness\\                                                                                                                                                                                                                                                                                                                                        \midrule
%Application & Relevant SIB & Relevant SIB & Fault-Related \\
%& & ($\textit{length} \geq 2$) & Method Calls\\

%\begin{tabular}[c]{@{}l@{}}Blocks with \\ RMs (\%)\end{tabular} & \begin{tabular}[c]{@{}l@{}}Average RMs\\ per Block\end{tabular} & \begin{tabular}[c]{@{}l@{}}Average \% of\\ RMs / MiBs\end{tabular} & \begin{tabular}[c]{@{}l@{}}Blocks with\\ RMs (\%)\end{tabular} & \begin{tabular}[c]{@{}l@{}}Average RMs\\ per block\end{tabular} & \begin{tabular}[c]{@{}l@{}}Average \% of\\ RMs / MiBs\end{tabular} \\ 

\midrule

BossTransfer & 50\%         & 75\%                                   & 77\%                \\
FakeGPS      & 60\%         & 86\%                                   & 79\%                \\
FilePicker   & 19\%         & 52\%                                   & 64\%                \\
GetBack GPS          & 45\%         & 50\%                                   & 40\%               \\
%GoodWeather  & 55\%         & 61\%                                   & 72\%               \\
GoodWeather  & 55\%         & 61\%                                   & 49\%               \\
KanjiFix     & 50\%         & -                                    & 30\%                  \\
MapDemo      & 63\%         & 69\%                                   & 45\%                 \\
%onInflate    & 39\%         & 50\%                                   & 36\%                 \\ 
PoGoIV & 24\% & 63 \% & 61\% \\
PrivacyPolice & 52\%			& 67\%					& 32\%\\	
QuotoGraph & 100\% & 100\% & 54\% \\
SearchView   & 48\%         & 50\%                                   & 26\%                \\ %\hline			\\
ToneDef   & 1\%         & 67\%                                   & 22\%                \\ %\hline			\\

\bottomrule
\end{tabular}
\end{footnotesize}
\end{center}
\vspace{-0.8cm}
\end{table}

We can notice that the density of SIBs that include fault-related method calls is rather sparse when considering the full set of blocks (column \emph{All SIBs}), \begin{empirical}ranging from 1\% to 100\%\end{empirical}. However, if we exclude the blocks with a single call, which are the blocks that contribute the least to our analysis, we can see that the density of blocks incorporating information about the fault increases significantly (column \emph{SIBs ($\textit{lenght} \geq 2$)}): \begin{empirical}at least half of the blocks are always relevant, with a density of relevant blocks reaching 86\% for FakeGPS and 100\% for QuotoGraph. There is an exception to this that is KanjiFix. In that case all the SIBs have length 1, thus the analysis can only be performed with the full set of SIBs. Note again that half of the blocks carry relevant information.\end{empirical}

The performance of the SIBs is quite variable in terms of their completeness. \begin{empirical}For some apps, such as BossTransfer, FakeGPS, FilePicker, and GoodWeather, the SIBs include most of the behaviors related to the fault, while for other apps, such as KanjiFix, SearchView, and ToneDef, the blocks are representative of about one fourth of the behavioral differences introduced with the fix.\end{empirical} It is important to emphasize two aspects. First, \techniqueName requires to capture some relevant differences in the executions but does not need to capture all the differences. \begin{empirical}In fact, as discussed in RQ2, \techniqueName managed to be  effective with KanjiFix where only 30\% of the relevant method calls have been captured.\end{empirical} Second, low percentages do not imply that \techniqueName is missing information present in the baseline and in the failure trace files (the diff procedure used in \techniqueName is essentially complete by construction), but typically correspond to completely new behaviors introduced with the fix that were not present in the previous version of the app.  

\smallskip 

In summary, although the information contained in the blocks is not noise-free, the blocks, especially the ones with non-negligible weights, are confirmed to carry relevant information about the failure and can be realistically exploited to identify the method responsible for the fault, as shown in RQ2.

\subsection{RQ2: \RQB}

The incompatibility problems introduced in our subject apps required the modification of a single method, with the exception of GoodWeather that required changing two methods. We do not consider the new methods introduced by developers to organize the code of the fix since these methods were not present in the faulty version of the app.

Table~\ref{tab:ranking} shows the results. Column \emph{\#SIB} indicates the number of SIBs exploited by \techniqueName in the localization, while column \emph{Pos Ranking} indicates the position of the method(s) that must be fixed in the ranking (we never experienced tie positions in the evaluation). Values below or equal to 5 are reported in bold.

\begin{table}[t]
\centering
\caption{Ranking.}
\label{tab:ranking}
\begin{footnotesize}
\begin{tabular}{@{}lcc@{}}
\toprule
Application      & \#SIB & Pos Ranking\\ \midrule
BossTransfer     & 4                      & \textbf{2}*                            \\
FakeGPS          & 21                     & \textbf{5}                            \\
FilePicker          & 29                     & \textbf{4}                            \\
GetBack GPS    & 16                     & 10*                            \\
GoodWeather    & 28                     & \textbf{1,2}                            \\
KanjiFix         & 6    			 & \textbf{1}           \\
MapDemo        & 13                     & 8                            \\ %\midrule
%onInflate        & 8                      & \textbf{1}                            \\
PoGoIV	&	16	&	7	\\
PrivacyPolice	&	12	&	\textbf{1}	\\
QuotoGraph	&	1	&	\textbf{1}	\\
SearchView        & 30                     & 9                            \\
ToneDef        & 3                     & -                            \\
%\textbf{Average} &       & 5.5         &                        & 4.285714286                  \\ 
\bottomrule
\end{tabular}
\end{footnotesize}
\vspace{-0.5cm}
\end{table}

The number of SIBs may vary significantly based on the failure. However, \begin{empirical}\techniqueName has been always able to report the method to be fixed within the top 10 positions, with the exception of ToneDef, where \techniqueName was unable to include the fixed method in the ranking, since the method was not part of the collected stack traces. The BossTransfer and GetBack GPS apps are considered special cases. In the former case the exact method to be fixed is not part of the failure call tree and thus the final ranking does not include that method. However, the method is in an anonymous class that is defined inside another method that occurs at the second place of the ranking. In the latter case the developers placed the fix in a method within an abstract class, whereas \techniqueName detected the override of the same method in the concrete class. Because of these special cases, we reported these results with a mark.\end{empirical}
%\review{6. Why isn't the method from the anonymous class part of the trace (Line 857)?
%\marco{Instrumenting anonymous classes with the Xposed module is tricky, since you should know the name of a class in order to attach hooks on its methods.}
%}
\begin{empirical}In seven cases the method to be changed was ranked among the top 5 positions with four perfect results, that is, the method to be changed is ranked at the top place.\end{empirical} 
This result is particularly good. In fact, practitioners have been reported to consider acceptable inspecting of up to 10 methods, with a preference for techniques that require inspecting 5 methods at most~\cite{Kochhar:LocalizationExpectation:ISSTA:2016}. %\techniqueName has always reported the target method in a position not higher than $8$, with two perfect outcomes.

%\marco{\techniqueName has always reported the target method in a position not higher than $10$, with three perfect outcomes.}

%\smallskip

%In summary, \techniqueName successfully localized several problems caused by framework upgrades.

\subsection{RQ3: \RQC}
%\review{
%for RQ3, I find it a bit artificial, as it basically ends up comparing apples and oranges. I understand that the authors may be concerned about not comparing with SBFL, and I appreciate their effort, but I am not sure that study adds much to the paper. An initial user study on the usefulness of the rankings produced by FILO, even if preliminary, would be a much more convincing study in my opinion.}
%
%\review{
%Comparison of the RQ3 is not fair in the sense that there is a lot of noise introduced by generating thousands of events (more than normally required) used as test cases for the approaches to compare with.}

This research question compares \techniqueName to alternative approaches that can be used to localize the method to be fixed. We identified two main alternatives. The first one is what we called naive trace analysis, that is, simply comparing the baseline and failure traces and inspecting anomalous methods calls in the order of occurrence. This approach is included to confirm the need of analyzing the failure in a more sophisticated way, as \techniqueName does, than simply comparing traces.

The second approach is  a classic SBFL method: Ochiai~\cite{Abreu:2007}. Although there are several alternative formulas that can be used~\cite{Wong:SurveyLocalization:TSE:2016}, we selected Ochiai because it is one of the most effective methods and a quick investigation based on alternative formulas has not revealed better results.

Note that SBFL techniques have stricter requirements than \techniqueName to be applied and produce a more limited output. In fact, they require a test suite with passing test cases to compute the ranking and do not provide any additional information that can help the developer interpreting the result produced by the technique. On the contrary, \techniqueName only requires a failing execution to be applied and augments the ranking with information about anomalous boundary calls that can help determining the reason of the failure. Since our apps are released without a test suite, in principle SBFL techniques would not be applicable to our cases. To overcome this issue, we generated a test suite of passing test cases with the Monkey automatic testing tool~\cite{Monkey}. We used a setting favouring the generation of a rather extensive test suite: we generate test cases by configuring Monkey to emit 10,000 events per test case, which is 200 times the default value, and we run Monkey for 10 minutes per app, which is double than the time that has been empirically reported to produce advances in the coverage~\cite{Choudhary:TestInputGeneration:ASE:2015}. We setup the testing environment to prevent the generation of failures and inspected the test execution to make sure that failing test cases have not incidentally included in the test suite. We collected coverage data at the level of both methods, to be consistent with the granularity of the other approaches, and statements, to investigate the effectiveness at a finer granularity. We then computed the ranking using Ochiai. 

Table~\ref{tab:comparison} reports the position of the method that must be modified to implement the fix in the ranking returned by \techniqueName, naive trace analysis, and Ochiai (since we have two methods to be modified, we have two positions for GoodWeather). Rows \emph{Top-1}, \emph{Top-5} and \emph{Top-10} indicate the number of times each technique has ranked the target method in the top, top 5 and top 10 positions, respectively. Row \emph{Not in the ranking} reports the number of times a technique has not included the target method in the ranking (marked with a ``-" in the respective row). For each row, the best result is shown in bold.

Naive trace analysis achieved the worst results. \begin{empirical}In six cases the method with the fix was not present in the ranking and in the other cases the number of methods to be inspected before reaching the target method was incredibly high, consisting of hundreds of entries in several cases.\end{empirical} This result confirms the unsuitability of simple trace analysis.

%In general, Ochiai performed sensibly worse than \techniqueName. 

\begin{empirical}Ochiai performed better than \techniqueName only twice, with the ToneDef and MapDemo apps. ToneDef escaped to \techniqueName, while could still be addressed with Ochiai.\end{empirical} The result obtained with MapDemo is due to the characteristic of the fault that is systematically activated every time a specific statement is executed. This case is favourable to SBFL techniques. 

%\begin{empirical}In the GoodWeather app, Ochiai (method) and \techniqueName performed the same. 
In all the other cases, \techniqueName outperformed Ochiai. In three cases, GetBack GPS, QuotoGraph, and SearchView, Ochiai (both methods and statements) could not generate any ranking since the apps failed during startup and it was impossible to produce a suite of passing test cases. On the contrary, \techniqueName effectively ranked the method to be fixed at the $10^{st}$, $1^{st}$ and $9^{th}$ position of the ranking.  In another case, FilePicker, Ochiai (statement) could not localize the fault because the fault is due to missing statements, while \techniqueName effectively ranked the target method at the $4^{th}$ position. In the remaining cases, \techniqueName always ranked the target method significantly better than Ochiai (both methods and statements). 

\begin{empirical}Overall, \techniqueName was not to able to rank the fix only once (1 out of 12). When successful, \techniqueName always ranked the target method below position 10 (11 out of 12 cases), ranking it at the top in four cases and in the top five positions in seven cases. Ochiai (method) failed to include the target method in the ranking three times, achieving a perfect ranking only twice, and missed to rank the target method in the top 10 positions in 9 out of 12 cases. Ochiai (statement) failed to include the right statement in the ranking in five cases, achieved only one perfect ranking, and missed to include the statement in the top 10 places in all the rest of the cases.\end{empirical}

Let us remark that \techniqueName is cheaper to execute than SBFL. In fact it can process traces in few seconds and its cost mainly depends on the execution of the failing test case. In contrast, SBFL techniques require the execution of complete test suites with programs instrumented to collect full coverage data, which is order of magnitude more expensive. 

\smallskip 

In summary, the results show that \techniqueName is more effective than naive trace analysis and SBFL with problems introduced by framework upgrades.

\begin{table}[th]
\caption{Comparison between \techniqueName, Naive trace analysis and Ochiai.} \label{tab:comparison}
\begin{footnotesize}
\begin{center}
\begin{tabular}{@{}lc c c c@{}}
\toprule
Application  & \techniqueName & Naive Trace & Ochiai & Ochiai \\
  &  &  Analysis & (method) & (statement) \\
\midrule

BossTransfer &  \textbf{2} & 138 & 4 & 32\\

FakeGPS      & \textbf{5} & 328 &  13 & 65\\

FilePicker   & \textbf{4} & 203 & 81 & -\\

GetBack GPS  & \textbf{10} & - & - & - \\

GoodWeather  & \textbf{1,2} & - & 1,32 & 5,5 \\

KanjiFix     & \textbf{1} & - & 19  & 23\\

MapDemo      & 8 & 45 & \textbf{1} & \textbf{1}\\

PoGoIV & \textbf{7} & - & 48 & 283  \\ %3

PrivacyPolice & \textbf{1} & 26 & 21 &130 \\

QuotoGraph & \textbf{1} & - & - &- \\

SearchView   & \textbf{9} & 109 & - & -\\

ToneDef   & - & - & 24 &  \textbf{13}\\ \midrule

Top-1 & \textbf{4} & 0 & 2 & 1\\
Top-5 & \textbf{7} & 0 & 3 & 2 \\
Top-10 & \textbf{11} & 0 & 3 & 2 \\
Not in the ranking & \textbf{1} & 6 & 3 & 4\\

%\emph{Average} & \emph{4.88} & \emph{141.5} &    \\ 
\bottomrule
\end{tabular}
\end{center}
\end{footnotesize}
\vspace{-0.5 cm}
\end{table}

\subsection{Discussion}
A characteristic of \techniqueName compared to other fault localization techniques is its capability to provide supporting evidence of the ranking in the form of anomalous calls that capture the symptoms of the failure. These symptoms can be conveniently used by the tester to better understand the failure and work on the implementation of a fix. In this section, we report qualitative results by discussing the output produced by \techniqueName for four of the faults analyzed in the empirical evaluation.

In GoodWeather the fix must be implemented in {\footnotesize \texttt{gpsRequestLocation}} and {\footnotesize \texttt{onOptionsItemSelected}}, which are the top ranked methods. This ranking is supported by two anomalous SIBs. Since each block is represented by the first call of the block, the information provided on the first place to the developer consists of two calls: a call to {\footnotesize \texttt{ContextCompat.checkSelfPermission}} and a call to {\footnotesize \texttt{LocationManager.requestLocationUpdates}}. The provided information clearly points at a permission issue (based on the anomalous call to {\footnotesize \texttt{checkSelfPermission}}) with the location services (based on the anomalous call to {\footnotesize \texttt{requestLocationUpdates}}). Notably the fix exactly consists of adding the code to grant the permission to access the location services from the {\footnotesize \texttt{gpsRequestLocation}} method. %, which is the top ranked method.

A similar case is MapDemo where \techniqueName associates the method that must be modified to implement the fix with an anomalous call to {\footnotesize \texttt{Location.getLatitude}}, which returns \texttt{null} because of the lack of permissions and the fix consists again of adding the code necessary to obtain this permission.

Another interesting case is PrivacyPolice. The fix is implemented in method {\footnotesize \texttt{ScanResultsChecker.onReceive}}, which is ranked at the top. The SIBs associated with these methods share the presence of the {\footnotesize \texttt{WifiManager.getScanResults}}, which is the framework method that experienced the behavioural change (always returns \texttt{null} if the GPS is disabled) and leads the app to a malfunction. The fix requires handling the result produced by this method properly.
%This ranking is enriched by a set of anomalous suspicious invocation blocks. While the representatives of each block, \texttt{WifiManager.disableNetwork} and \texttt{Builder.build} are not directly relevant to the bug, they both contains invocation to \texttt{WifiManager.getScanResults} which is the framework method that experienced the behavioural change (always returns \texttt{null} if the GPS is disabled) and leads the app to a malfunction.

A particular case is represented by QuotoGraph, which crashes on startup without producing a trace. 
The result is the whole baseline trace treated as a single block of suspicious invocations clearly suggesting problems with the initialization of the app. \techniqueName correctly pointed at the {\footnotesize \texttt{LWQApplication.onCreate}} method, which in fact has been fixed modifying the initialization procedure of the app.

%\marco{A particular case is represented by QuotoGraph. The bugged version of the app crashes at the onCreate of the application, which leads to an emtpy bug trace. The result is the whole baseline trace treated as single block of suspicious invocations in which the fix locus is not present. The ranking obtained from the callgraph, however, points correctly to the }

%\marco{In PrivacyPolice the fix must occur in \texttt{ScanResultsChecker.onReceive}, which is the top ranked method returned by \techniqueName. This ranking is enriched by a set of anomalous suspicious invocation blocks. While the representatives of each block, \texttt{WifiManager.disableNetwork} and \texttt{Builder.build} are not directly relevant to the bug, they both contains invocation to \texttt{WifiManager.getScanResults} which is the framework method that experienced the behavioural change (always returns \texttt{null} if the GPS is disabled) and leads the app to a malfunction.}

%MapDemo: il fix va in com.example.mapdemo.MapDemoActivity\$2.onClick, l'anomalia che lo evidenzia è android.location.Location.getLatitude. Questo è il metodo che viene invocato e scatena la nullPointerException ed il crash dell'applicazione perché l'oggetto Location su cui lo invoca è null a causa della mancanza di permessi per ottenere la posizione corrente tramite l'invocazione a com.google.android.gms.location.getLastLocation().
%\leo{Abbiamo bisogno almeno di un caso non relativo ai permessi da discutere qui}
 
\subsection{Threats to Validity}
There are two main threats affecting the validity of our evaluation. One is an internal validity threat and is about the comparison to Ochiai. As discussed, \techniqueName only requires a failing execution to be applied, while Ochiai, and other SBFL techniques, needs a test suite of passing test cases to be applied, and such a test suite was not available for the apps used in our evaluation. This case is quite frequent in practice, for instance the vast majority of the apps in GitHub are developed without having an associated test suite. This practical scenario confirms the value of \techniqueName being independent on suites of passing test cases. To anyway obtain information about the comparison of \techniqueName to SBFL we derived a quite extensive test suite of passing test cases for each app and then applied SBFL. In principle, since the outcome of the localization depends on the test suite, we cannot know if the results would be different using another test suite. However, we worked conservatively generating as many passing test cases as possible to obtain the best localization from Ochiai, so this is unlikely to happen. 

The second threat is an external validity threat and concerns the generalization of the reported evidence. %We selected applications on GitHub based on an objective procedure to avoid any bias. %The size of the evaluation does not allow to fully generalize the results, but 
\techniqueName performed consistently good in the studied incompatibilities and the steps of the analysis are based on general concepts, without including any ad-hoc optimization. This provides a good degree of confidence on the general validity of the results. %suggests that the results reported in the paper might be generally valid.

\subsection{Limitations}

\begin{empirical}Although \techniqueName performed well with almost all the subjects,\end{empirical} there are cases that cannot or can be extremely hard to address with our technique. We discuss three relevant cases below.% They are quite unlikely to happen in practice, and this is why we have not experienced them in our evaluation. We however discuss these cases here for the sake of completeness.

\emph{Faulty method outside the set of collected stack traces}: In principle a faulty method might be missing from all the collected stack trace instances. \techniqueName collects a number of these instances, one for each SIB, thus it is unlikely that the faulty method falls outside every stack trace instance. \begin{empirical}In our evaluation happened only in 1 case out of 12.% (see  the fault in the ToneDef app).
\end{empirical}

\emph{Faulty configuration}: In some cases the fix might require changing a configuration file of the app rather than changing the app. These faults are outside the scope of \techniqueName. %, which targets faults in the app. 

\emph{New callback methods}: Problems caused by new callback methods, not existing in the base environment, cannot be detected by \techniqueName. In some cases, these problems might be however introduced together with other faulty changes that \techniqueName can detect, such as in the case of GoodWeather.

\section{Related Work}
\label{sec:relatedWork}

Framework upgrades are frequent in mobile ecosystems and Android in particular~\cite{McDonnell:APIStability:ICSM:2013,AndroidVersionHistory}. Many of these upgrades are intentionally \emph{not} backward compatible and app developers struggle adapting their apps to newer versions of the framework, as witness by the many discussions opened on Stack Overflow every time a framework upgrade is released~\cite{Linares-Vasquez:StackOverflowDiscussions:ICPC:2014}. When these upgrades are not handled properly, apps might be affected by fragmentation-induced compatibility issues~\cite{Wei:AndroidFragmentation:ASE:2016}, as well as maintenance and security issues~\cite{Li:DeprecatedAPI:MSR:2018}.

\smallskip

Many techniques focused on the \emph{detection of the incompatibilities} introduced with these upgrades. For instance, CiD~\cite{Li:CompatibilityIssues:ISSTA:2018} builds API lifecycle models based on  the revision history of the Android framework and uses these models to detect incompatibilities. In their work Mostafa et al. \cite{Mostafa:BackwardIncompatibilities:ISSTA:2017} detect behavioral backward incompatibilities in Java libraries, including Android, by runnning regression tests and checking bug reports. Similarly Mora et al.~\cite{Mora:ClientSpecificEquivalenceChecking:ASE:2018} defined an approach based on the lazy exploration of the behavioral space to assess if a library update may impact on the clients. DiffDroid detects inconsistent app behaviors across devices~\cite{Fazzini:DiffDroid:ASE:2017}.
All these approaches focus on the \emph{detection of misbehaviours}. On the contrary, \techniqueName identifies \emph{the place where the fix should be implemented} and \emph{the symptoms} are used to assist the work of the developers who investigate the failures.  

%. In the second module, an analysis is performed to locate and extract the Android API usage pattern in an application. This module considers not only the main code of the application, but also any code to be dynamically uploaded available in the application package. Finally, the analysis preserves a summary of the conditions under which the extracted APIs can be achieved. Finally, the third module comprehensively evaluates the results of the previous modules to report any potential API compatibility issues.

%Mostafa et al. \cite{Mostafa:BackwardIncompatibilities:ISSTA:2017} detected behavioral backward incompatibilities in the Java libraries, including Android, by runnning regression tests for version pairs and checking bug reports for version updates. The majority of behavioral backward incompatibilities are not well documented in API documents or release notes. GUI behavior change is one of the major cause of client bugs related with behavioral backward incompatibilities.

%Wei et al. \cite{Wei:AndroidFragmentation:ASE:2016} have focused on fragmentation-induced compatibility issues They observed that 35\% were non-device-specific issues and their occurrences were due to Android platform evolution. These issues are independent from device models and can affect a wide range of device models running specific Android OS versions.

Some techniques can be used to reduce the compatibility problems. For instance,  ReBA~\cite{Dig:ReBA:ICSE:2008} is a technique for the development of libraries augmented with adapters to guarantee backward compatibility. While this might be an option for the developers who want to put extra effort on releasing backward compatible components, in many practical cases developers intentionally release upgrades that are not backward compatible.

\smallskip 

% Fault localization
Instead of detecting failure symptoms, fault localization techniques can be used to attempt to localize faults. A number of them use the coverage profile of the passing and failing test cases to perform the localization. These techniques are usually referred to as \emph{spectrum-based fault localization} (SBFL)~\cite{Wong:SurveyLocalization:TSE:2016}. Notable examples are Tarantula~\cite{Jones:Tarantula:ASE:2005}, Ochiai~\cite{Abreu:2007}, and Zoltar~\cite{Riboira:GZoltar:TAICPART:2010}. These techniques can be potentially used to address a broader class of faults than \techniqueName, in fact they are not limited to the case of framework upgrades. On the other hand, \techniqueName is more effective than SBFL both in terms of its applicability, indeed \techniqueName \emph{does not require a test suite of passing test cases}, and its effectiveness, since it produces \emph{failure symptoms} and \emph{a more accurate localization}, based on our evaluation. Furthermore, SBFL has shown limitations~\cite{Pearson:EvaluatingSBFL:ICSE:2017,Steimann:ThreatsSBFL:ISSTA:2013,Parnin:debugging:ISSTA:2011} and can hardly satisfy the requirements that a practical fault localization approach should satisfy based on the opinion of practitioners~\cite{Kochhar:LocalizationExpectation:ISSTA:2016}. On the contrary, \techniqueName can be executed quickly, can produce accurate rankings, and can isolate symptoms that help understanding the failure and interpreting the ranking, which is a key requirement already highlighted in empirical studies~\cite{Parnin:debugging:ISSTA:2011}. Finally, API documentation could also be exploited to facilitate the API migration process~\cite{Lamothe:APIMigration:MSR:2018}.  

\smallskip
% Dynamic analysis
Finally, \techniqueName exploits \emph{anomaly detection} in the analysis~\cite{Chandola:AnomalyDetectionSurvey:ACMCS:2009}. There exist a number of solutions that analyze and compare the behaviors of programs and components to identify anomalies that can be used to support the debugging activity. For instance, BCT~\cite{Mariani:BCT:TSE:2011}, RADAR~\cite{Pastore:Radar:ISSRE:2012}, and the technique by Pradel and Gross~\cite{Pradel:SpecMining:ICSE:2012} perform this kind of analysis in the context of component-based systems, regression testing, and object-oriented software, respectively. Mimic performs a similar analysis in the attempt to analyze reproduced field failures~\cite{Zuddas:Mimic:ASE:2014}. Differently from these techniques, \techniqueName originally combines fault-localization and anomaly detection, exploiting anomalies to both perform the localization and augment the ranking with symptomatic information about the failure.

\section{Conclusions}
\label{sec:conclusions}

Timely fixing problems caused by framework upgrades is important to make mobile apps compatible with the latest releases of the operative systems. %Since discovering the source of incompatibilities and identifying the changes to be implemented in the app can be time-consuming, developers may exploit localization techniques~\cite{Wong:SurveyLocalization:TSE:2016,Jones:Tarantula:ASE:2005,Abreu:2007} to obtain recommendations on the faulty methods that must be changed to fix the problem. Unfortunately, localization techniques are often imprecise, require the availability and the execution of full test suites to be applied, and do not provide any supporting information beyond the mere localization of a method.  
\techniqueName can assist developers when performing this task by automatically identifying the faulty method that must be fixed to solve the compatibility issue, and reporting selected anomalous events observed in the failing execution, to facilitate the analysis of the problem. %In line with expectation of practitioners~\cite{Kochhar:LocalizationExpectation:ISSTA:2016}, \techniqueName suggests the location of the fix and provides hints about the reason of the failure, with an analysis that requires few minutes only to be completed.\textbf{Verify if hints can be really generated}

The evaluation shows that \techniqueName can be accurate and practical. %~\cite{Kochhar:LocalizationExpectation:ISSTA:2016}. 
Moreover it has weaker requirements and higher effectiveness than SBFL in the domain of faults caused by framework upgrades. As part of future work, we intend to investigate the possibility of applying \techniqueName to other contexts, such as  library evolution, to extend our approach with automatic program repair capabilities~\cite{Gazzola:Repair:TSE:2017}, and to experiment with fixes that span multiple methods and files.
% to other operating systems for mobile devices, and extend the approach to other contexts that imply software evolution. 

\begin{small}
\subsubsection*{Acknowledgements}
This work has been partially supported by the H2020 ERC CoG Learn project (grant agreement n. 646867).
\end{small}

\bibliographystyle{IEEEtran}
\bibliography{Main.bib}

\end{document}